\documentclass[12pt]{iopart}
\usepackage[pdftex]{graphicx}

\begin{document}

\title[]{Critical tunneling currents in the regime of bilayer excitons}

\author{L. Tiemann, W. Dietsche, M. Hauser and K. von Klitzing}
\address{Max-Planck-Institut f\"{u}r Festk\"{o}rperforschung, Heisenbergstra{\ss}e 1, 70569 Stuttgart, Germany}
\ead{L.Tiemann@fkf.mpg.de}

\begin{abstract}
We have investigated the tunneling properties of an electron double
quantum well system where the lowest Landau level of each quantum
well is half filled. This system is expected to be a Bose condensate
of excitons. Our four-terminal dc measurements reveal a nearly
vanishing interlayer voltage and the existence of critical tunneling
currents $I_{critical}$ which depend on the strength of the
condensate state.
\end{abstract}

\maketitle

\section{Introduction}

Macroscopic quantum systems such as superconductors and superfluids
are the remarkable consequence of many of bosonic particles
occupying the same lowest energy state, and thus forming a
Bose-Einstein condensate (BEC). The design of closely spaced two
dimensional electron systems (2DES) which can be contacted
independently is the foundation to create a BEC of excitons in
semiconductors \cite{Fertig1989, MacDonald1994}. Exposed to a strong
perpendicular magnetic field $B$, the density of states of each of
the 2DES will condense into a discrete set of sub-bands, the Landau
levels. The total number of occupied states is then parameterized by
the filling factor $\nu=\frac{hn}{eB}$. If the electron densities
$n$ are tuned to be identical in both layers, the filling factors
will simultaneously be at $1/2$ at a particular $B$. Governed by
Coulomb interactions, the bilayer system can then be viewed as a
Bose condensate of interlayer quasi-excitons by coupling an electron
from layer 1 to a vacant state from layer 2 and vice versa. Since
these excitons have an infinite life time, their properties can be
investigated via electrical transport experiments.

Transport experiments in the counter-flow configuration
\cite{Tutuc04, Kellogg04}, where constant currents of equal
magnitude but opposite direction are imposed on the two layers have
indeed shown that exclusively if $\nu_{layer~1}+\nu_{layer~2}\approx
1$ (denoted as ''total filling factor 1'', or simply
''$\nu_{tot}=1$''), the Hall and longitudinal voltages across both
layers (nearly) vanish. While this by itself can be interpreted as
the result of a dissipationless flow of charge-neutral electron-hole
pairs in one direction, interlayer tunneling experiments
\cite{Champagne07, Spielman04, Spielman01, Spielman00} have shown an
I/V characteristic that has an astonishing resemblance to the one of
the Josephson effect. However, the bilayer at $\nu_{tot}=1$ is only
partially analogous to a Josephson junction \cite{Rossi2005}, and it
is important to recognize the experiment as tunneling between two
electron systems that {\textit{only as a whole}} form the correlated
state \cite{Park2006}. This fact might also explain why no true dc
supercurrent at zero bias has been observed so far. Suitable bilayer
samples are required to be weakly tunneling \cite{Spielman00},
however, they only possess a very small single electron tunnel
splitting $\Delta_{S,AS}$ of up to approximately 100~$\mu$K. Even
though interlayer phase coherence is completely \emph{spontaneous}
only for $\Delta_{S,AS}\rightarrow 0$, it has been demonstrated
\cite{Murphy1994} that single electron tunneling can co-exists with
this correlated state which is still dominated by Coulomb
interactions.

Our interlayer tunneling experiments indicate that the Bose
condensation strongly changes the nature of the tunneling process.
More specifically, we exploit a pure dc tunneling configuration
which reveals the existence of critical tunneling currents
$I_{critical}$. These critical currents terminate the regime of
interlayer phase coherence, i.e., when the total current $I$ exceeds
the threshold value of $I_{critical}$, the 4-terminal interlayer
resistance abruptly increases by many orders of magnitude.

\section{Samples}

Our data originate from three different samples from the same wafer.
The double quantum well structure consists of two 19~nm GaAs quantum
wells, separated by a 9.9 nm superlattice barrier composed of
alternating layers of AlAs (1.70 nm) and GaAs (0.28 nm). The quantum
wells have an intrinsic electron density of about $4.5\times
10^{14}$~m$^{-2}$ and a low-temperature mobility which exceeds
40~m$^2$/Vs. While sample A is a standard Hall bar geometry with a
length of 880~$\mu$m and a width of 80~$\mu$m, samples B and C are
patterned into a quasi-Corbino ring \cite{Tiemann08}, both with an
outer diameter of 860~$\mu$m and a ring width of 270~$\mu$m. A
commonly used selective depletion technique \cite{Eisenstein90,
Rubel97} was used to provide separate contacts to the layers. The
densities in the two layers are balanced with a front and back gate
which cover the entire region of the structures including the edges.

\section{Tunneling Setup}

The modulation of a tunable dc bias $V_{dc}$ with a low amplitude ac
sine wave $V_{ac}$ which is applied between the two layers (i.e.,
the interlayer bias) is a convenient and commonly used method to
determine the differential conductance $dI_{ac}/dV_{ac}$. While a
$V_{dc}\neq 0$ counter-shifts the Fermi energies of both systems,
$V_{ac}$ is used to induce an ac (tunneling) current which can be
detected via a sensitive lock-in technique. In the zero magnetic
field case, if both layers have identical densities and
$V_{dc}\approx 0$, the Fermi energies of both layers align, and
owing to momentum and energy conservation, electron tunneling
becomes possible. Under the application of a magnetic field,
however, it generally requires a finite energy e$V_{dc}$ to
add/extract an electron to/from one of the correlated 2DES
\cite{Eisenstein91}. This means that no peak in $dI/dV$ centered
around $V_{dc}$=0 is expected under application of a (strong)
perpendicular magnetic field.

\section{Experimental data}

\subsection{AC Modulation of a DC Interlayer Bias}
\begin{figure}[!htp]
\centering
 \includegraphics[width=0.8\textwidth]{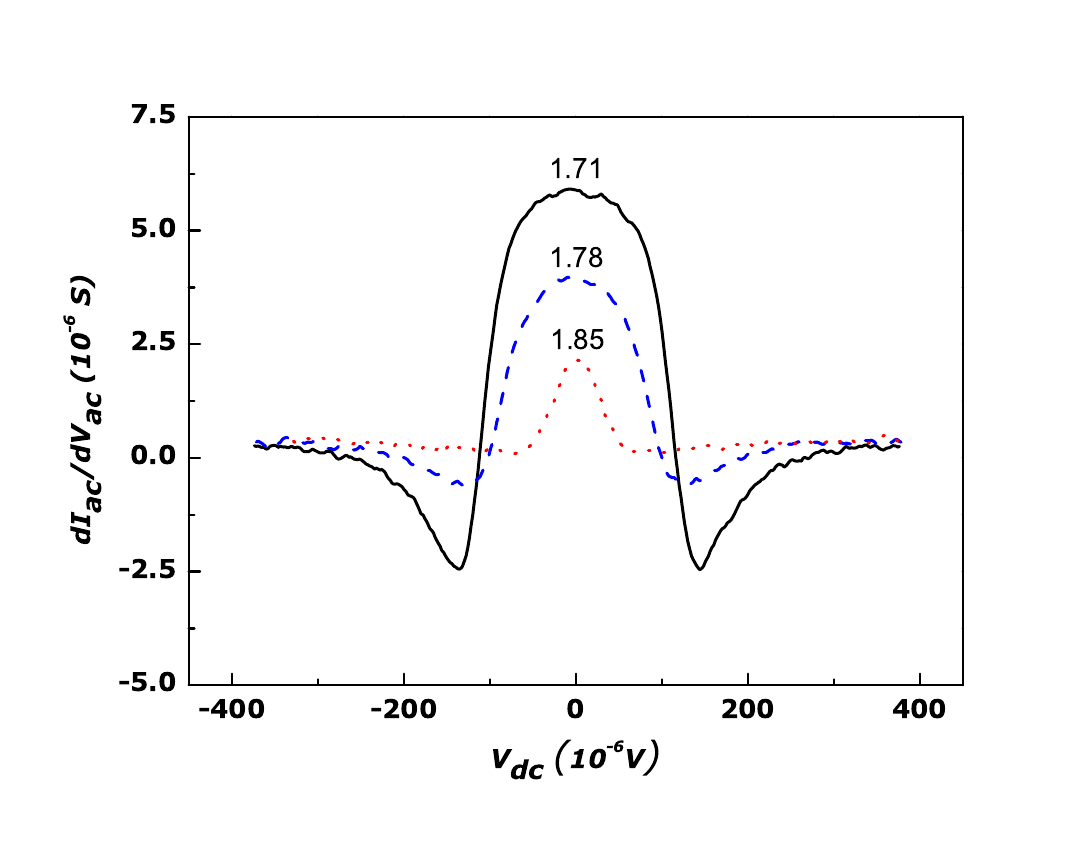}
 \caption{Differential tunnel conductance for $d/l_B=1.85$~(dotted line),~$1.78$~(dashed line),~$1.71$~(solid
 line). In addition to its amplitude, also the width increases with
 decreasing $d/l_B$. These data were produced on sample A at $T_{bath}\approx 25$~mK.}
 \label{fig:1}
\end{figure}

Figure \ref{fig:1} shows the results of the common tunneling
experiment as previously described. The tunable dc bias was
modulated with a small ($\approx 7~\mu$V) ac voltage. The current
was detected by measuring the voltage drop across a 10~k$\Omega$
resistor connected towards common ground. These measurements were
performed on sample A (Hall bar) at $T_{bath}\approx 25$~mK and
$\nu_{tot}=1$ with balanced carrier densities in the two layers
leading to three different $d/l_B=\{1.85, 1.78, 1.71\}$. This ratio
of the center-to-center distance $d$ between the layers (here
$28.9$~nm) and the magnetic length $l_B=\sqrt{\hbar /eB}$
characterizes the strength of the $\nu_{tot}=1$ state due to Coulomb
interactions.

For Figure \ref{fig:1} we use the common notation where we plot the
2-point (2pt) differential conductance $dI/dV$ versus the 2pt
voltage $V_{dc}$, i.e., the curve illustrates the measured $dI_{ac}$
induced by the ac modulation of 7~$\mu$V versus the variable dc
interlayer bias. The peaks centering $V_{dc}=0$ can be identified as
the familiar enhanced tunneling anomaly \cite{Spielman00} of the
$\nu_{tot}=1$ state. From high to low values of $d/l_B$, the full
width at half maximum (FWHM) in these three cases is about
60~$\mu$V,~160~$\mu$V and 200~$\mu$V. While an increase of the
tunneling amplitude upon decreasing $d/l_B$ was to be expected based
on earlier reports, it is yet remarkable that its FWHM appears to
increase as well. As we will show next, the answer to this apparent
inconsistency is hidden in the modality of a two-terminal tunneling
experiment where the interlayer resistance becomes much smaller than
other series resistances.

\subsection{Pure DC Interlayer Tunneling}

\begin{figure}[!htp]
\centering
 \includegraphics[width=0.85\textwidth]{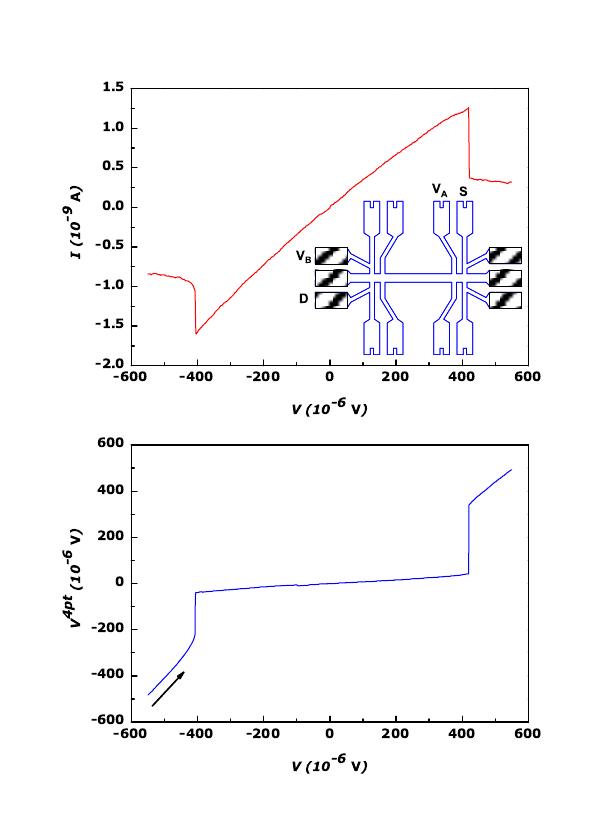}
 \caption{Top panel: measured current plotted versus the 2pt interlayer bias $V$. Clearly visible
 are critical currents below which the characteristic has a steeper slope. Bottom panel: 4pt voltage
 $V^{4pt}$ which was measured simultaneously versus $V$. A plateau exists around zero bias where the 4pt voltages is
 nearly zero. The plateau terminates at exactly the same 2pt voltages where the critical
 currents occur. The inset shows a schematics of the experiment with the source (S) and drain (D)
 contacts and the location of the voltage probes $V_{A,B}$. Shaded contacts connect to the lower layer.
 These data were produced on sample A at a $d/l_B$ of 1.44 and $T_{bath}\approx 25$~mK.}
 \label{fig:2}
\end{figure}

Using a sufficiently sensitive dc measurement setup, the tunneling
experiment can be simplified by measuring the dc current directly.
In addition, with a separate pair of contacts, a 4-point (4pt) setup
is possible to probe the dc voltage that drops across the barrier as
well. Figure \ref{fig:2} illustrates these 4pt measurements,
performed again on sample A at $\nu_{tot}=1$ for a single $d/l_B$ of
1.44. The current was again detected by measuring the voltage drop
across a resistor connected towards the common ground. The top panel
thus illustrates this dc current as a function of the 2pt dc
voltage. Consistent with the prior observation of an enhanced
tunneling conductance at small bias voltages, the dc current
displays a relatively steep slope around $V_{dc}=0$ which abruptly
terminates when the current exceeds values of approximately -1.5~nA
or +1.25~nA \footnote{The very sharp jump of the current and voltage
at about 400~$\mu$V in Figure \ref{fig:2} cannot be directly
compared to the results in Figure \ref{fig:1} which are measured at
a different value of $d/l_B$. In addition, we observed a hysteretic
behavior which may lead to a smearing of the curves in ac modulated
measurement.}. The existence of such a critical current had already
been predicted \cite{Rossi2005,Park2006} but had not been clearly
demonstrated. Most strikingly, the 4pt measurements on the bottom
panel reveal a plateau in the probed dc voltage close to zero which
accompanies the region where the current flow is enhanced.

We emphasize that the unexpected increase in width of the
differential tunneling conductance curves in Figure \ref{fig:1} is
now explainable in terms of a strongly reduced 4pt dc voltage. In
view of this reduction, which affects the 4pt ac modulation as well,
the $dI/dV$ curves would rescale to a very narrow peak with a very
high amplitude if plotted versus $V^{4pt}$ and with
$dV=V^{4pt}_{ac}$. Please note that at no other (total) filling
factor is such a behavior observable, i.e., the strong reduction of
the 4pt interlayer voltages is a peculiarity of the $\nu_{tot}=1$
state.

We associate the range in which $V^{4pt}$ is small with a state in
which interlayer coherence is established. Its width apparently
depends on the sum of all series resistances $R_s$ in the system,
such as contact arms and the series resistance for the current
measurement. If $R_s$ is large compared to the 4pt interlayer
resistance, the experiment is essentially performed by controlling
the current so that the existence of critical currents can be
resolved. In a voltage-controlled experiment on the other hand,
critical currents are concealed in the limit of vanishing
(interlayer) resistances.

\begin{figure}[!htp]
\centering
 \includegraphics[width=0.8\textwidth]{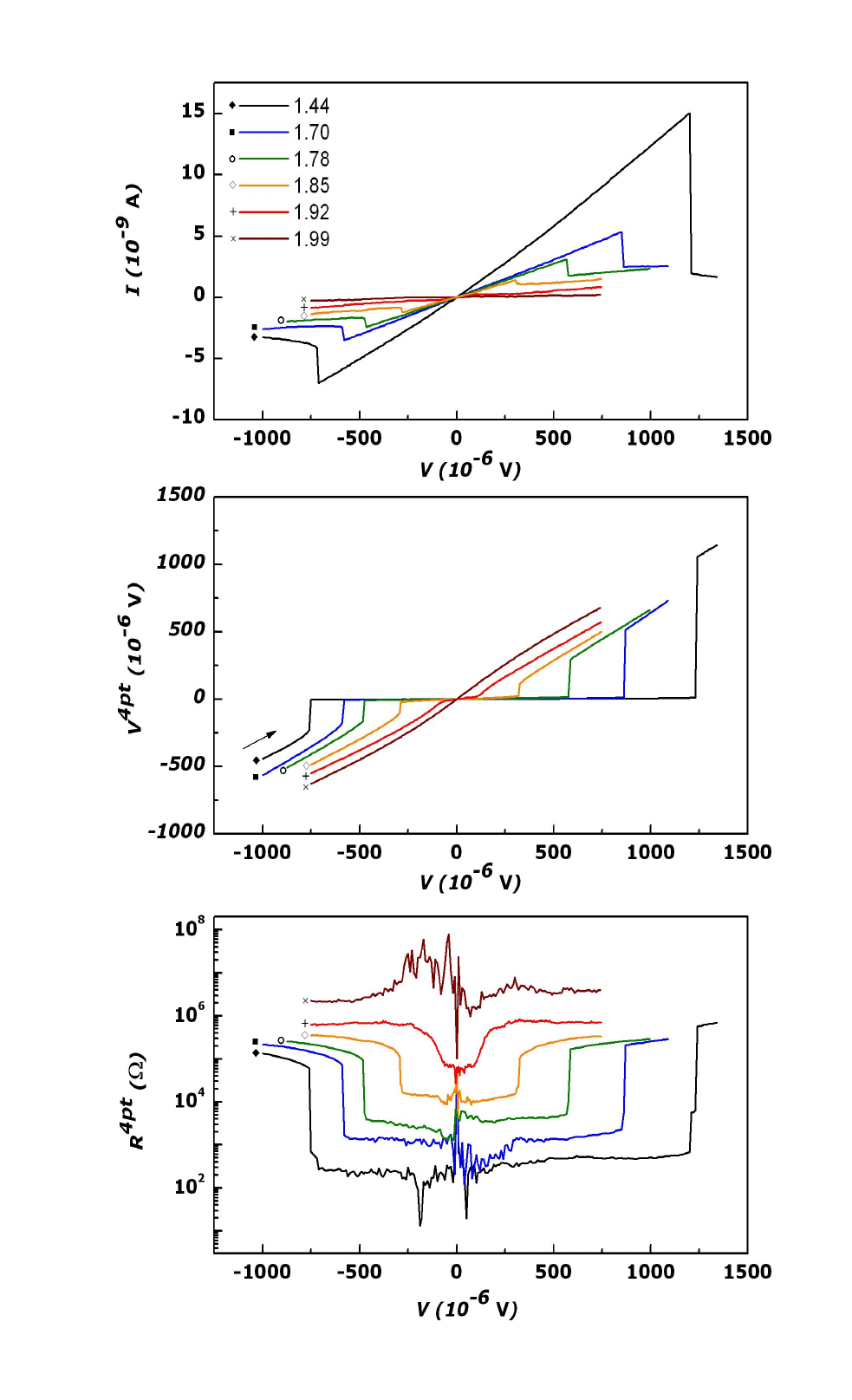}
 \caption{The top panel plots the measured tunneling current versus the 2pt
 interlayer bias $V$ for a set of six different $d/l_B$=\{1.99, 1.92, 1.85, 1.78, 1.70, 1.44\}.
 The mid panel shows the probed 4pt voltage $V^{4pt}$ which was not measured
 simultaneously, and the bottom panel illustrates the calculated 4pt interlayer
 resistance. The enhanced noise around $V=0$ originates from the
 noise in detecting small voltages.}
 \label{fig:3}
\end{figure}

Figure \ref{fig:3} demonstrates how the tunneling process evolves
upon reducing the ratio $d/l_B$ from high to low values, i.e., upon
reducing the electron densities in both layers simultaneously and
adjusting the magnetic field. These data were produced on sample B
(Corbino geometry) where the voltage $V$ was applied between the two
outer circumferences of the upper and lower layer. Moving from high
to low values of $d/l_B$, plateaus in the 4pt voltage appear which
progressively take on lower values. At the same time the critical
currents grow. The resulting 4pt interlayer resistance has, at the
lowest $d/l_B$, a value of only about 200~Ohms at $T_{bath}\approx
25$~mK. Once the critical current is exceeded, the 4pt interlayer
resistance is nearly of the same magnitude for all $d/l_B$, which
suggests that the condensate is destroyed and the current is
maintained by bare electron tunneling. The observed asymmetry which
is particularly pronounced in sample B for low $d/l_B$ is owing to a
strong hysteresis.

\begin{figure}[!htp]
\centering
 \includegraphics[width=0.8\textwidth]{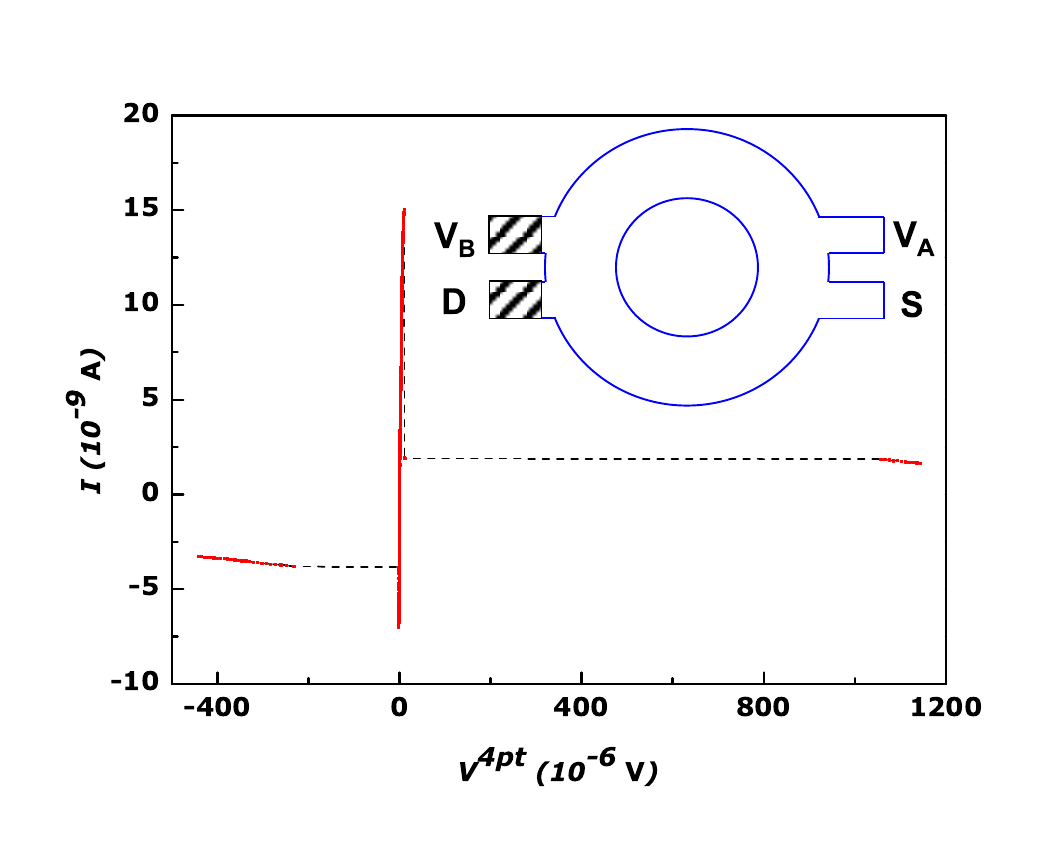}
 \caption{The current plotted versus the measured 4pt voltage $V^{4pt}$ for $d/l_B=1.44$ (sample B). Only red dots
  are actual data points, the black dashed lines are used to guide the eye. The inset shows a simplified schematics
  of the experiment where the contacts to probe the voltage and the source and drain contacts are marked. Shaded
  contacts connect to the lower layer. Unused Ohmic contacts are disregarded.}
 \label{fig:4}
\end{figure}

The question of the lowest obtainable 4pt resistance and/or its
accuracy is directly related to the question which factors influence
the 4pt voltage. In addition to a temperature-activated behavior, it
is relevant where exactly the potential is probed (see insets in
Figures \ref{fig:2} and \ref{fig:4}) because residual resistances
come into play. More precisely, \emph{any} current $I$ that crosses
the boundary of a two-dimensional electron system under quantum Hall
conditions will produce a voltage difference across the contact of
the order of the Hall voltage h/e$^2\cdot I$ ($\approx$25~$\mu$V at
$I=10^{-9}$~A). Since the sign of the Hall voltage depends on the
sign of the magnetic field $B$, it should be possible to account for
its influence by inverting the magnetic field. And indeed, the
inversion from $+B_{\nu_{tot}=1}$ to $-B_{\nu_{tot}=1}$ also
inverted the slope of $V^{4pt}$ around $V=0$ in Figure \ref{fig:2}.
The mean value calculated from the curves at $+B$ and $-B$, however,
did not completely cancel out $V^{4pt}$ within the plateau region.
This might be caused by longitudinal resistance components, if the
current flows through dissipative regions \cite{Fertig2005}.
Nevertheless, as we have shown, this (residual) voltage and the
resulting 4pt interlayer resistance was a lot smaller for sample B
(Corbino). For this sample, the voltage was probed in a
"longitudinal" configuration, i.e., the voltage was probed (across
the barrier) at contacts that lie between the source and drain and
at the same side of the current flow.

Figure \ref{fig:4} finally shows the current plotted versus the 4pt
voltage for $d/l_B=1.44$. In this representation our data resemble
earlier reports \cite{Spielman01} where however the maximal current
was of order 20~pA, or about 1000 times smaller. Note that in
\cite{Spielman01} the I/V characteristic was deduced from
integrating the differential tunneling conductance data which may
have masked the critical current behavior reported here. Even though
the sample characteristics differ only marginally (QW/barrier/QW
width in \cite{Spielman01} is 18~nm/9.9~nm/18~nm) which yields a
comparable value of $\Delta_{S,AS}$, the effective single particle
tunneling amplitude in our samples appears to be larger. Hence, we
assume that the different magnitudes of the maximal currents can be
attributed to a different bare interlayer tunneling which strongly
influences the tunneling anomaly at $\nu_{tot}=1$ \cite{Rossi2005}.

For reasons of completeness, we would like to elaborate on an
experimental detail. Generally, the application of an interlayer
bias will imbalance the electron densities of both layers, while the
total density $n_{tot}=n_{layer~1}+n_{layer~2}$ remains constant.
This has the consequence that the regular quantum Hall (QH) states
will shift to lower/higher fields, owing to a higher/lower density
in the respective single layer. The $\nu_{tot}=1$ QH state on the
other hand depends only on $n_{tot}$ and thus does not shift to a
different magnetic field. Using the Shubnikov-De Haas oscillations
in transport experiments in the low field regime, we were able to
adjust front and back gate voltages while sweeping the interlayer
bias to keep the density in each of the two layers constant.
However, interlayer tunneling experiments did not significantly
differ from unadjusted measurements. The bias-induced imbalance for
the electron bilayer system in question is $\approx 4~\%$ for
$V^{2pt}$=500~$\mu$V, but might become irrelevant as the (effective)
4pt interlayer bias nearly vanishes for $I<I_{critical}$.

\subsection{Parallel Tunneling: ''The Load Configuration''}

In a different experiment, sample C (Corbino ring) was set up in a
drag experiment as described in \cite{Tiemann08}, where a voltage is
applied across only one layer (drive layer), while the other (drag)
layer is kept as an open circuit. Only at a total filling factor of
one has this been shown to produce a voltage drop of equal sign and
magnitude across the adjacent drag layer. At the same time, the
conductance through the drive layer vanishes, i.e., the bulk is in a
gapped state.

\begin{figure}[tp]
\centering
 \includegraphics[width=0.9\textwidth]{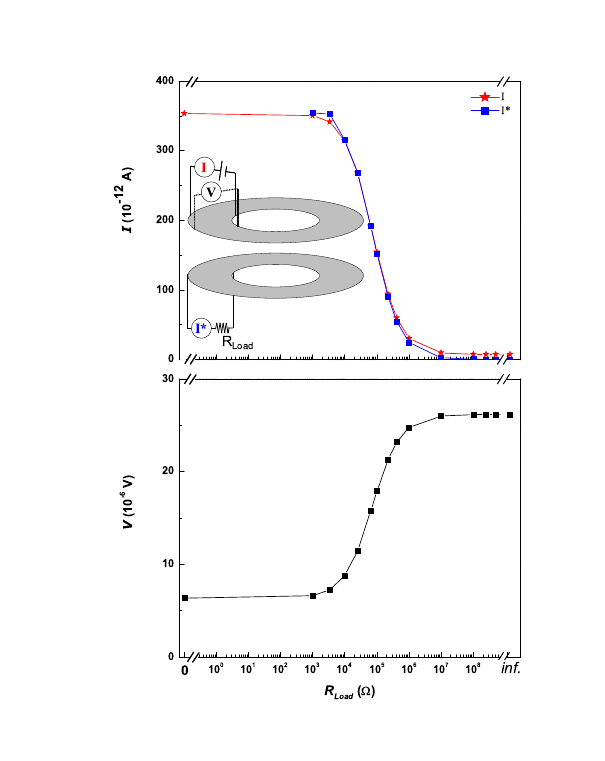}
 \caption{Results of the parallel-tunneling configuration. Top panel: currents measured in drive
 and drag circuit as a (log) function of the load resistance $R_{Load}$. As soon as $R_{Load}$
 is sufficiently reduced, the drive current $I$ increases, even though the bulk is in a gapped
 state. Meanwhile, a current $I^*$ with the same sign and magnitude through the load resistance
 can be measured. The inset shows a schematics of the experiment in a pseudo-3d view. Bottom panel:
 4pt drive voltage $V$ as a function of $R_{Load}$. These data were produced on sample C at a
 $d/l_B=1.60$ and $T_{bath}\approx 14$~mK.}
 \label{fig:5}
\end{figure}

In this situation we applied a variable resistor $R_{Load}$ between
the inner and outer circumference of the drag layer as shown in the
inset of Figure \ref{fig:5}. For $R_{load}\rightarrow\infty$ the
system behaves as before. However, upon decreasing $R_{Load}$ from
$\infty$ to $0$ Ohms, a current begins to flow through the bridge
connecting inner and outer circumference of the drag layer.
Simultaneously, a current of equal magnitude in the circuit of the
drive layer can be measured. In the light of the strongly reduced
4pt interlayer resistance we demonstrated with Figure \ref{fig:2}
through \ref{fig:4}, these results can be explained in terms of
parallel-tunneling at both sample edges carried by quasiparticles
\cite{MacDonald07, MacDonald08}.

The observables in any of these transport geometries are the
currents and voltages \emph{in the leads} \cite{MacDonald07}, so we
have no direct access to what is happening within the bulk. The
ground state of our bilayer system in the correlated regime can be
described by the Halperin (111) state \cite{Halperin1983}, as the
Laughlin wave function describes the ground state of the fractional
quantum Hall effect (FQHE). And like in the FQHE, it is convenient
to introduce the \emph{quasiparticle} concept. These quasiparticles
experience enhanced interlayer tunneling, just as the quasiparticle
Hamiltonian of a superconductor has pair creation and annihilation
terms. The quasiparticles in our system arise at the interface where
the single particle electron current from the leads meets the
correlated $\nu_{tot}=1$ phase. Since for $\nu_{tot}=1$ the bulk of
the drive layer is in a gapped state, it prohibits any regular
single electron current flow across the annulus. However, in a
process which is analogous to Andreev reflection \cite{MacDonald08},
the injection of a single electron leads to a condensate current, or
an excited state from the condensate ground state, respectively. To
put it simply, every incident single electron in the top layer
excites an exciton in the bulk. To conserve total charge in both
layers, or to counter for this sudden net flow of excitons in the
bulk, respectively, an electron must exist into the leads in the
bottom layer. The Bose condensate thus changes the single electrons
into quasiparticles which are easily transferred. This constant flow
of quasiparticles is the process we would like to refer to as
\emph{quasiparticle tunneling} \cite{MacDonald07}.

If the inner and outer circumference of the bottom layer are
physically connected over a sufficiently small resistance
$R_{Load}$, it offers a short cut path across the gapped bulk for
reflected single electrons. Once having passed that bridge, each
electron will by itself undergo the same process of triggering
condensate currents and quasiparticle tunneling at the other edge.
While this model is able to account for our data, we cannot
definitively say whether this configuration really allows us to
trigger such an excitonic current through the bulk of the
$\nu_{tot}=1$ QH state or not. It is also possible that some still
unknown (tunneling) process is taking place.

\section{Summary and Conclusion}

We have presented dc tunneling experiments on electron double layer
systems at a total filling factor of one which clearly show the
existence of critical tunneling currents $I_{critical}$. When the
total current $I$ exceeds the critical value, the 4pt interlayer
resistance increases by many orders of magnitude. The results can be
explained in terms of quasiparticle tunneling which is possible due
to the Bose condensation. These observations could have grave
consequences for the interpretation of the $\nu_{tot}=1$ QH state
and the transport experiment performed within this regime. This
could be of particular relevance if the currents that are imposed in
regular transport are smaller than $I_{critical}$.

\section{Acknowledgments}

We thank J. G. S. Lok for the design of the Corbino geometry and J.
H. Smet for giving us access to some of his equipment. Also, we
would like to acknowledge the German Ministry of Research and
Education (BMBF) for its financial support and gratefully thank both
Allan H. MacDonald and Ady Stern for discussions.

\end{document}